\documentclass[prl,preprint]{revtex4}
\usepackage{pifont}
\usepackage[dvips]{graphicx}
\usepackage{latexsym}
\usepackage{verbatim}
\usepackage[T1]{fontenc}
\usepackage{textcomp}
\def \be{\begin{equation}}
\def \ende{\end{equation}}

\usepackage{xspace,amssymb,amsfonts,amsmath}
\usepackage{color}

\begin{document}

\title{Evidence for Klein tunneling in graphene \emph{p-n} junctions}%
\author{N. Stander}
\author{B. Huard}
\author{D. Goldhaber-Gordon}
\email[Corresponding author~: ]{goldhaber-gordon@stanford.edu}
\affiliation{Stanford University, Department of Physics, Stanford, California, USA}%

\begin{abstract}

Transport through potential barriers in graphene is investigated
using a set of metallic gates capacitively coupled to graphene to modulate the potential landscape. When
a gate-induced potential step is steep enough, disorder becomes less important and the
resistance across the step is in quantitative agreement with predictions
of Klein tunneling of Dirac fermions up to a small correction. We also perform
magnetoresistance measurements at low magnetic fields and compare
them to recent predictions.
\end{abstract}
\maketitle

Graphene is promising for novel applications 
and fundamental physics due to its remarkable electronic, optical and
mechanical properties~\cite{neto_electronic_2007}. At energies
relevant to electrical transport, quasi-particles are believed to behave like Dirac
fermions with a constant velocity $v_F \simeq 1.1\times
10^6~\mathrm{m.s}^{-1}$ characterizing their dispersion relation
$E=\hbar v_Fk$. The Klein paradox for massless Dirac fermions
predicts that carriers in graphene hitting a potential step at
normal incidence transmit with probability one regardless of the
height and width of the step~\cite{katsnelson_chiral_2006}. At
non-normal incidence, this tunneling problem for 2D massless fermions can be represented as a 
1D problem for massive Dirac fermions, with the effective mass proportional to the conserved transverse momentum. The Klein tunneling probability should then depend on the profile of the potential step
\cite{katsnelson_chiral_2006,cheianov_selective_2006,beenakkerreview}.
Recent experiments have investigated transport across potential
steps imposed by a set of electrostatic gates
\cite{huard_transport_2007,marcus,kim_npn,delft,Gorbachev_air_bridge,lau} and results of Ref.~\cite{Gorbachev_air_bridge} support an interpretation of Klein tunneling. 
We present measurements on six devices which allow a
quantitative comparison with Klein tunneling in graphene when the potential profile created by the gates is evaluated realistically~\cite{zhang_nonlinear_2007}. Disorder is
sufficiently strong in all our devices to mask effects of multiple
reflections between the two steps of a potential barrier, so that
all data can be accounted for by considering two independent steps
adding ohmically in series. Finally, we probe the transition from clean to disordered transport across a single potential step, and we refine 
the accuracy of the transition parameter introduced by Fogler \emph{et al.}
\cite{fogler_effect_2008}. In a complementary measurement, we show
that the effect of a low magnetic field on the Klein tunneling across a potential step in graphene is not explained by existing predictions in the clean
limit~\cite{Levitov}.
\begin{figure}[hbtp]
\begin{center}
\includegraphics[width=8cm]{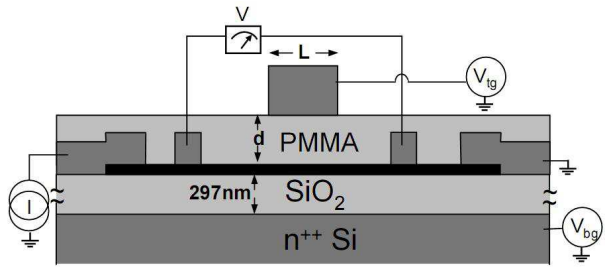}
\caption{Schematic diagram of a top-gated graphene device with a 4-probe measurement setup. Graphene sheet is black, metal contacts and gates dark grey. \label{scheme}}
\end{center}
\end{figure}
\begin{table}[hbtp]
\begin{center}
\begin{tabular}{|c|c|c|c|c|c|}\hline
Sample & $L$ (nm) & $w$ ($\mu$m) & $d$ (nm) &
\hspace{0.1cm}$\left\langle \beta\right\rangle$\hspace{0.1cm} & $\mu$ ($\mathrm{cm}^{2}\mathrm{V}^{-1}\mathrm{s}^{-1}$)
\\\hline
A60 & 60 & 4.3 & 34 & 7.6 & 1800\\
B100 & 100 & 2.1 & 42 &  3.8 & 1700\\
B220 & 220 & 2.1  & 42 & 3.5 & 1700\\
C540 & 540 & 1.74  & 25 & 7.9 & 1400\\
A860 & 860 & 3.6 & 34 & 7.9 & 1800\\
C1700 & 1700 & 1.74 & 47 & 1.9 & 1300\\
\hline
\end{tabular}
\caption{Geometrical properties of the samples: top gate length $L$,  graphene strip width (interface length) $w$, and  top gate dielectric
thickness $d$. Same letter for two device labels indicates same graphene
sheet. All dimensions were taken by both Scanning Electron Microscopy and Atomic Force Microscopy. The transition parameter $\beta$ between clean and diffusive transport in a single \emph{p-n} junction is also shown (see text), averaged across the whole measured voltage range such that $n_{\mathrm{bg}}<0$ and $n_{\mathrm{tg}}>0$. Counter-intuitively, despite devices' low mobility, $\beta\gg 1$ so that Klein tunneling is expected rather than diffusion across the interface. \label{Table1}}
\end{center}
\end{table}
\begin{figure}[hbtp]
\begin{center}
\begin{minipage}{.49\linewidth}
\includegraphics[height=7cm]{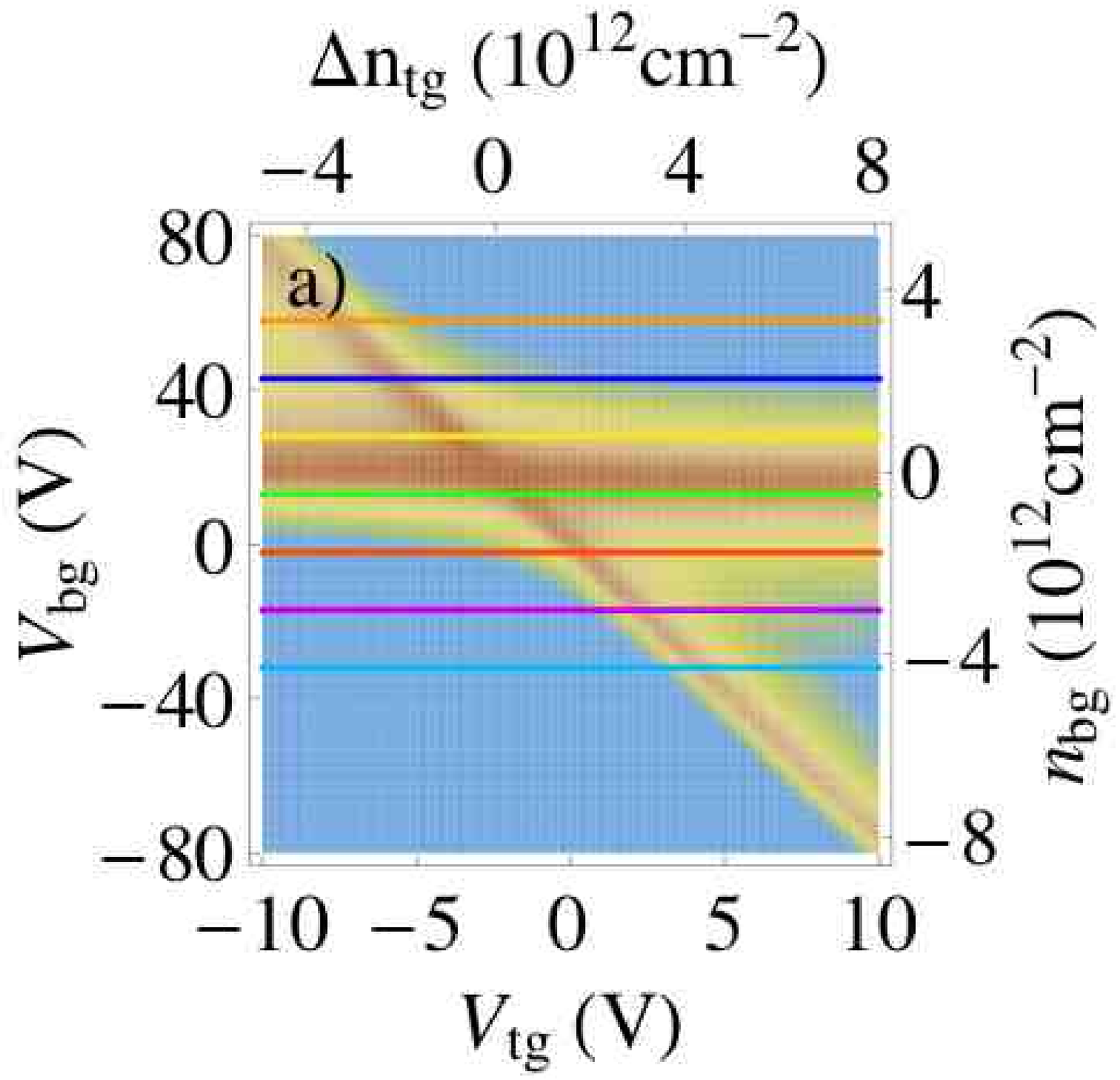}
\end{minipage} \hfill
\begin{minipage}{.49\linewidth}
\includegraphics[height=7cm]{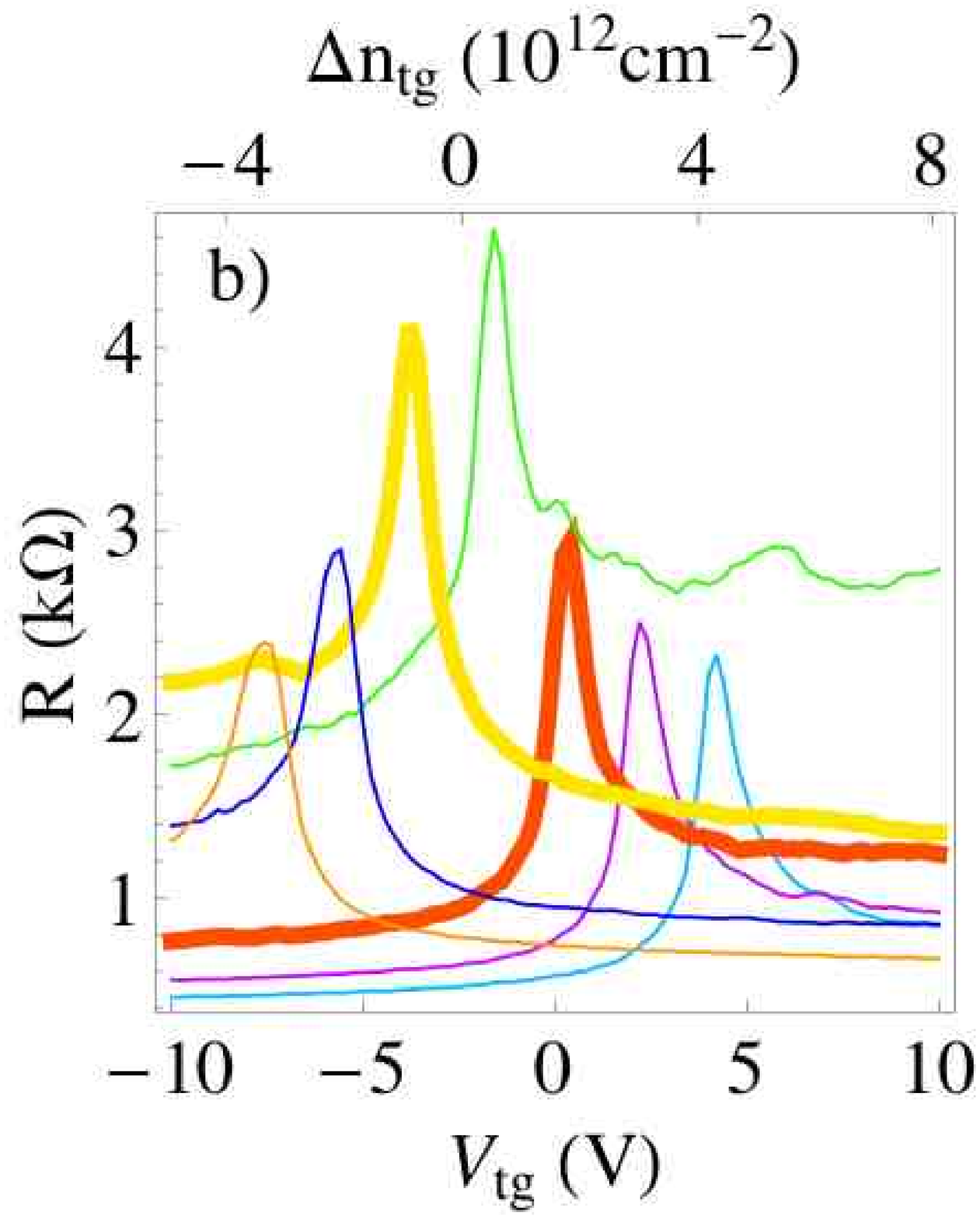}
\end{minipage} \hfill
\caption{\textbf{a)} 4-probe resistance measured on device C540 (see
Table ~\ref{Table1}), as a function of $V_{\mathrm{bg}}$ and
$V_{\mathrm{tg}}$. The color scale can be inferred from the cuts
shown in b. The densities $n_{\mathrm{bg}}$ and $\Delta
n_{\mathrm{tg}}$ are estimated using
$V_{\mathrm{tg}}^{0}=2.42~\mathrm{V}$,
$V_{\mathrm{bg}}^{0}=18.65~\mathrm{V}$ and
$C_{\mathrm{tg}}=107~\mathrm{nF.cm}^{-2}$. \textbf{b)} Resistance as
a function of $V_{\mathrm{tg}}$ at several values of
$V_{\mathrm{bg}}$. The two bold curves show a clear asymmetry with
respect to the peak ($n_{\mathrm{tg}}=0$) for both
$V_{\mathrm{bg}}<V_{\mathrm{bg}}^{0}$ (red) and
$V_{\mathrm{bg}}>V_{\mathrm{bg}}^{0}$ (yellow).\label{Fig1}}
\end{center}
\end{figure}

We measure six top-gated graphene devices (typical schematic shown in 
Fig.~\ref{scheme}), whose essential parameters are listed in Table~\ref{Table1}.  The
density $n_{\mathrm{bg}}$ far from the top-gated region is set by
the back gate according to
$n_{\mathrm{bg}}=\frac{C_{\mathrm{bg}}(V_{\mathrm{bg}}-V_{\mathrm{bg}}^{0})}{e}$
where $C_{\mathrm{bg}}=13.6~\mathrm{nF.cm}^{-2}$ is the back gate
capacitance per area (from Hall effect measurements on a similar
wafer oxidized in the same furnace run), $e$ is the electron charge, and
$V_{\mathrm{bg}}^{0}$ is the gate voltage required to attain zero
average density~\cite{Fuhrer1}. The density $n_{\mathrm{tg}}$ well
inside the top gated region is set by both back gate and top gate
voltages according to
$n_{\mathrm{tg}}=n_{\mathrm{bg}}+\frac{C_{\mathrm{tg}}(V_{\mathrm{tg}}-V_{\mathrm{tg}}^{0})}{e}$,
where $C_{\mathrm{tg}}$ and $V_{\mathrm{tg}}^{0}$ are the top gate
counterparts of $C_{\mathrm{bg}}$ and $V_{\mathrm{bg}}^{0}$.
Throughout this letter we use the notation $\Delta
n_{\mathrm{tg}}=n_{\mathrm{tg}}-n_{\mathrm{bg}}$ to identify the
contribution of the top gate voltage only, which tunes the potential
step height. As described in previous work
~\cite{huard_transport_2007}, an asymmetry with respect to
$n_{\mathrm{tg}}=0$ appears in the 4-probe resistance measured
across a top-gated region as a function of $V_{\mathrm{tg}}$ for
fixed back gate voltages $V_{\mathrm{bg}}$ (Fig.~\ref{Fig1}b). This
asymmetry quantifies the resistance across the potential step in
graphene created by the gates. All graphene top-gated devices were fabricated in the same way, which is described in detail in the Supporting material~\cite{EPAPS}.
For electrical characterization, samples are immersed in liquid Helium at 4~K and four-terminal
measurements are made using a lock-in amplifier at a frequency 32~Hz
with a bias current of 100~nA. All samples show typical monolayer graphene
spectra measured by Raman spectroscopy and exhibit the quantum Hall
plateaus characteristic of graphene when measured in
perpendicular magnetic fields up to 8~T at 4~K (see Supporting
material~\cite{EPAPS}).

In order to extract the resistance of the \emph{p-n} interfaces
only, we measure the odd part of resistance $R_{\mathrm{odd}}$ about
$n_\mathrm{tg}=0$~\cite{huard_transport_2007}:
\begin{equation}
2R_{\mathrm{odd}}(n_{\mathrm{bg}},n_{\mathrm{tg}})\equiv R(n_{\mathrm{bg}},n_{\mathrm{tg}})-R(n_{\mathrm{bg}},-n_{\mathrm{tg}}),
\label{Rodd}
\end{equation}
where $R$ is the four-terminal resistance as a function of the
densities far from the top gated region and well inside that region.
Extracting the odd part $R_{\mathrm{odd}}$ from the measured
resistance requires an accurate determination of the densities
$n_{\mathrm{bg}}$ and $n_{\mathrm{tg}}$. This is made by the
measurement of three independent quantities $V_{\mathrm{bg}}^{0}$,
$V_{\mathrm{tg}}^{0}$, and $C_{\mathrm{tg}}/C_{\mathrm{bg}}$. We
carefully measure these quantities by using the quantum Hall
measurements at 8~T and electron-hole symmetry\cite{EPAPS}. There
are two physical interpretations for $R_{\mathrm{odd}}$ depending on
the relative magnitude of two length scales: the mean free path
$l_{e}=\frac{h}{e^{2}}\frac{\sigma}{2\sqrt{\pi n}}$ (well defined
for $k_F l_{e}\gg 1$ or equivalently for a conductivity $\sigma \gg 2e^{2}/h$)
and the top gate length $L$. For $L\gg l_{e}$, after crossing the
first interface of the barrier carriers lose all momentum
information before impinging on the second interface. In this case,
the total barrier resistance can be modeled by two junctions in series. The expression $2(R_{\mathrm{pn}}-R_{\mathrm{pp}} )$ where $R_{\mathrm{pn}}$ ($R_{\mathrm{pp}}$) denotes the theoretical value of the resistance of a single \emph{p-n} ( \emph{p-p}) interface, can then be compared directly to the experimental quantity $2R_{\mathrm{odd}}$~\cite{huard_transport_2007}.   
For $ L\ll l_{e}$, multiple reflections occur between the two interfaces of the barrier, which is predicted to reduce the total barrier resistance~\cite{EPAPS}.    
\begin{figure}[htbp]
\begin{center}
\begin{minipage}{.49\linewidth}
\includegraphics[width=8cm]{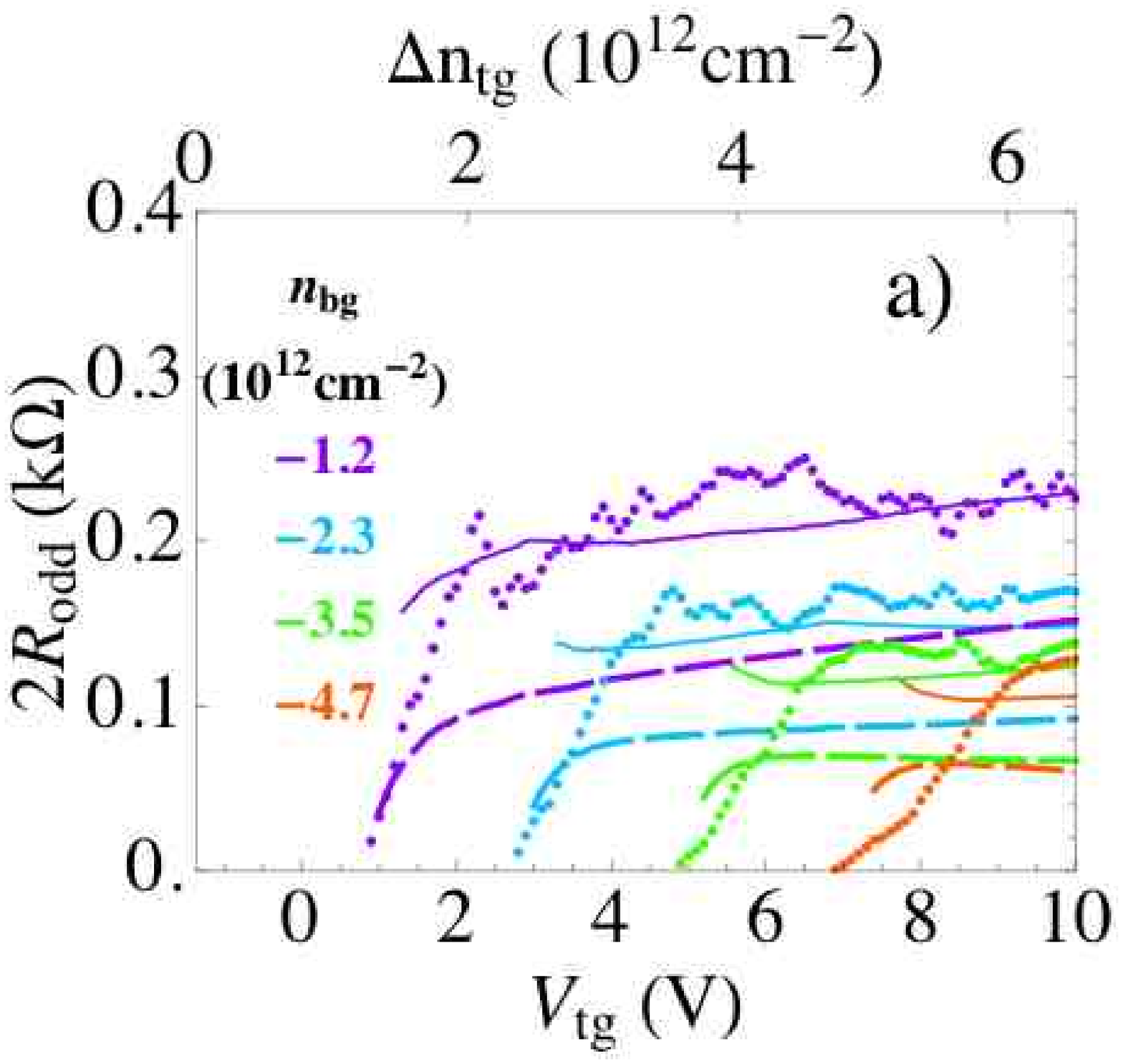}
\end{minipage} \hfill
\begin{minipage}{.49\linewidth}
\includegraphics[width=8cm]{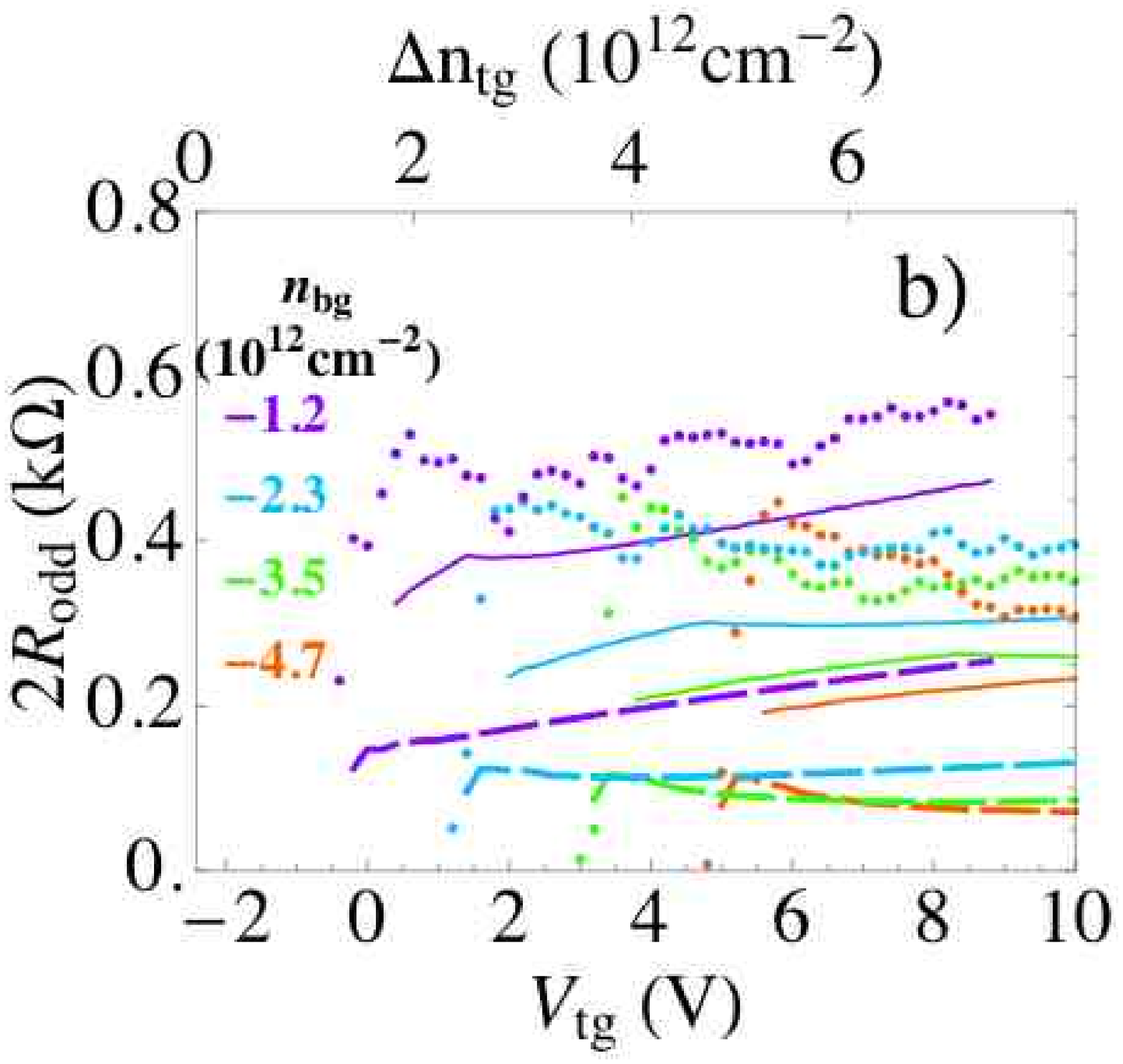}
\end{minipage} \hfill
\caption{\textbf{a)} The series resistance $2R_{\mathrm{odd}}$ of
the barrier interfaces as a function of $V_{\mathrm{tg}}$, for
several values of  $V_{\mathrm{bg}}$ for device A60 (corresponding
densities $n_{\mathrm{bg}}$ are labeled). The measured resistance
$2R_{\mathrm{odd}}$ (dots) is compared to the predicted value
$2(R_{\mathrm{pn}}-R_{\mathrm{pp}})$ using either a diffusive model, Eq.~(\ref{Rdif}) (dashed lines)  or a ballistic model Eq.~(\ref{Rall}) with the value $c_1=1.35$ chosen to best fit all six devices. (solid lines).
\textbf{b)} Same as a) for device C540. \label{Fig3}}
\end{center}
\end{figure}
As all devices have modest mobility, we start by using a diffusive model to calculate $R_{\mathrm{pn}}$ and $R_{\mathrm{pp}}$. In this model, due to disorder the resistance depends on the local resistivity $\rho(n)$ (measured for a uniform density at $V_{\mathrm{tg}}=V_{\mathrm{tg}}^{0}$) at each position $x$:
\begin{equation}
 R_{\mathrm{pn}}^{(\mathrm{dif})}-R_{\mathrm{pp}}^{(\mathrm{dif})} =\frac{1}{w}\int \rho(n(n_{\mathrm{bg}},n_{\mathrm{tg}},x))-\rho(n(n_{\mathrm{bg}},-n_{\mathrm{tg}},x))\mathrm{d}x
\label{Rdif}
\end{equation}
Figure~\ref{Fig3} compares the experimental curves for $2R_{\mathrm{odd}}$ as a function of
$V_{\mathrm{tg}}$ at several $V_{\mathrm{bg}}$ for samples A60 and C540 to the corresponding predictions. Clearly, the diffusive model represented by the dashed lines predicts resistance values considerably below the experimental curves, hinting that transport through the device cannot be viewed as entirely diffusive. 
Following the calculation by Fogler \emph{et al.}~\cite{fogler_effect_2008}, 
we retain the diffusive model for the region away from the interface, but replace it by a ballistic interface model for a region extending one mean free path in either direction from the location where density changes polarity~\cite{footnote_xbal_le}. Thus, 
\begin{equation}
R_{\mathrm{pn}}-R_{\mathrm{pp}}=R_{\mathrm{pn}}^{(\mathrm{bal})}- R_{\mathrm{pp}}^{(\mathrm{bal})}+R_{\mathrm{pn}}^{(\mathrm{dif})}|_{x\ge |l_e|}-R_{\mathrm{pp}}^{(\mathrm{dif})}|_{x\ge |l_e|}
\label{Rall}
\end{equation}
where the two last terms are taken from Eq.~\ref{Rdif}, but with the integral excluding  $x \in [-l_e,l_e]$. 
The first two terms are the ballistic contributions to the interface resistance for bipolar and monopolar configurations, and can be calculated individually  as follows.  All conduction channels on  the low-density side of a monopolar junction should have transmission nearly 1 through the junction~\cite{Cayssol}, so $R_{\mathrm{pp}}^{(\mathrm{bal})}=\frac{4e^2}{h} \frac{w \sqrt{\pi min(|n_{\mathrm{bg}}|,|n_{\mathrm{tg}}|))}}{2\pi}$. The bipolar case was addressed
by Zhang and Fogler \cite{zhang_nonlinear_2007}:
\begin{equation}
R_{\mathrm{pn}}^{(\mathrm{bal})} = c_1 \frac{h}{e^{2}w}
\alpha^{-1/6} |n'|^{-1/3}, \label{Rnp}
\end{equation}
where $h$ is Planck's constant, $\alpha=\frac{e^{2}}{\epsilon_r
\hbar v_{F}} \sim 0.56$ is the dimensionless strength of Coulomb
interactions ($\epsilon_r\approx 3.9$ is the average dielectric
constant of SiO$_{2}$ and cross-linked PMMA measured at 4K), and
$n'$ is the slope of the density profile at the position where the
density crosses zero (density profile calculated from the
\emph{classical Poisson equation with realistic gate geometry, temporarily treating graphene as a perfect conductor}). 
Expression~\ref{Rnp} refines this calculation to take into account non-linear screening of graphene close to zero density, going beyond the linear model used in Ref.~\cite{cheianov_selective_2006}.  
The prefactor $c_1$ in Eq.~(\ref{Rnp}) is determined numerically~\cite{zhang_nonlinear_2007}. In our case, $\alpha=0.56$ and the prefactor is predicted to be $c_1=1.10\pm 0.03$~\cite{fogler_private}. In order to test this prediction $c_1$ will be used as a single fit parameter across all samples and densities. The solid lines in Fig~\ref{Fig3} were generated by Eq.~\ref{Rall}, choosing $c_1=1.35$ to best account for all experimental curves in all devices (voltages
$V_{\mathrm{bg}}>V_{\mathrm{bg}}^{0}$ give a similar agreement, not
shown for clarity). The slight discrepancy between theoretical and experimental values of $c_1$ might be due in part to exchange and correlation effects. 
Trying to fit the data using a naive linear potential model requires an independent fitting parameter for each device, and even with the best fit to the data, some qualitative trends of the experimental data cannot be accounted for by this model, as described in detail in the Supporting material \cite{EPAPS}. This mismatch between the linear model and the data indicates the importance of accounting for non-linear screening close to zero average density.
We continue by calculating the ratio
$\eta=R_{\mathrm{odd}}/\left(R_{\mathrm{pn}}-R_{\mathrm{pp}}\right)$,
for all devices, for all measured $V_{\mathrm{bg}}$ and $V_{\mathrm{tg}}$, using Eq.~\ref{Rall}. The
histogram of $\eta$ is sharply peaked at a certain value
$\eta_{\mathrm{peak}}$ with a small peak width ~\cite{EPAPS}.
For all devices except C1700, regardless of their length $L$, $\eta$ is close or slightly higher than 1 when using $c_1=1.35$ (Fig.~\ref{Fig4}), which indicates that the resistances of both interfaces of the potential barrier simply add in series, and a single \emph{p-n} junction is less
sensitive to disorder than transport between the two interfaces of a
potential barrier. Fogler \emph{et al.} introduced the parameter
$\beta=n' n_{i}^{-\frac{3}{2}}$ to describe the 
clean/disordered transition in a single \emph{p-n} junction, where $n_i$ is
related to the mobility by $n_i=\frac{e}{\mu
h}$~\cite{fogler_effect_2008}. According to
Ref.~\cite{fogler_effect_2008}, when $\beta\gg 1$ the ballistic contribution in Eq.~(\ref{Rall}) dominates and the junction is in the clean limit, whereas for $\beta\ll 1$, the diffusive contribution in Eq.~(\ref{Rall}) dominates and the junction is in the disordered limit. The threshold $\beta = 1$ marks the transition where ballistic contribution must be taken into account since it is comparable to the diffusive contribution. 
In the following, we refine this transition threshold experimentally.
From Fig.~\ref{Fig4} and Table~1, it seems that transport is indeed
well described by Eq.~(\ref{Rall}) when $\beta>3.5$ but more poorly
for C1700 where $\beta\lesssim 2$, where we find that $\eta$ is further than 1 and has a large spread of values. In addition, Fogler \emph{et al.} predict that the diffusive contribution to the the interface resistance will be negligible for $\beta>10$, which is reached in several of our devices for densities $n_{tg}>3\cdot10^{12}cm^{-2}$. At these densities, in spite of our devices' modest mobility, the junction can be considered as disorder-free since the calculated ballistic contribution to $R_{\mathrm{odd}}$ is 10 times higher than the diffusive one, which allows us to make a rather accurate measurement of the ballistic contribution alone in this clean limit, and match it well with the ballistic terms in Eq.~(\ref{Rall}). 
\begin{figure}[hbtp]
\begin{center}
\includegraphics[height=8cm]{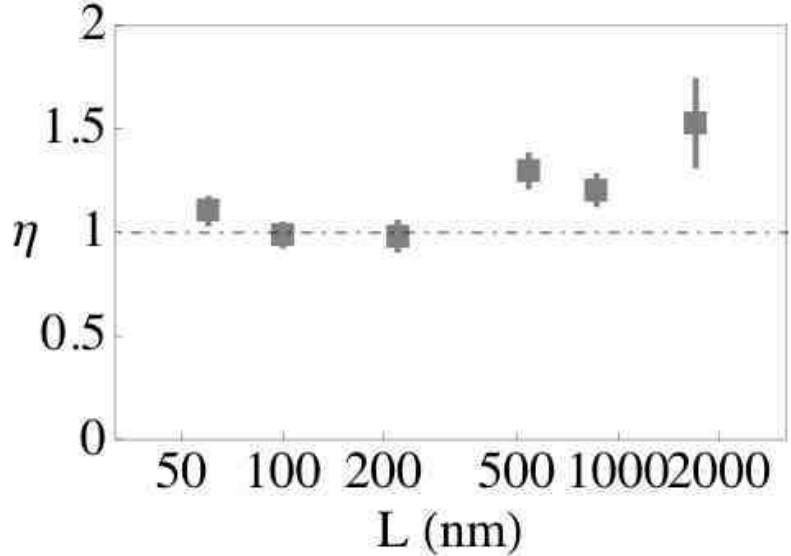}
\caption{Symbols: ratio
$\eta=R_{\mathrm{odd}}/\left(R_{\mathrm{pn}}-R_{\mathrm{pp}}\right)$
as a function of top gate length $L$ for the devices of Table~1. $R_{\mathrm{pn}}$ is calculated with $c_1=1.35$. The
vertical lines show the width of the histogram of $\eta$ for densities such that $\left|n_{\mathrm{bg}}\right|$,$\left|n_{\mathrm{tg}}\right|>10^{12}\mathrm{cm}^{-2}$.
 The dashed line at $\eta=1$ corresponds
to perfect agreement between theory and experiment, in the case
where the total resistance is the sum of the resistances of two
\emph{p-n} interfaces in series. \label{Fig4}}
\end{center}
\end{figure}
In a recent experiment where suspended top gates were used, 
for one sample the agreement with Eq.~\ref{Rdif}  -- the disordered limit -- was very good (sample S3 in Ref. [9]). This is due to a much larger distance between the top
gate and the graphene sheet, and much smaller density range than in the present work, likely due to lower dielectric constant combined with mechanical instability of the top gate when applying higher voltages. These two factors considerably
reduce $n'$ (around 80 times), which is not fully balanced by the cleaner
graphene of
Ref.~\cite{Gorbachev_air_bridge} ($n_i$ 2-5 times smaller). We estimate $\langle
\beta\rangle\approx 0.7$ for device S3 reported in
Ref.~\cite{Gorbachev_air_bridge}. Note that two other devices on substantially  cleaner graphene  (S1 and S2 in Ref.~\cite{Gorbachev_air_bridge}) support an interpretation of Klein tunneling with $\beta=2.5$ and $\beta=4$ respectively. From the present work and
from the result of Ref.~\cite{Gorbachev_air_bridge}, one can see
that the transition between clean and disordered transport in
\emph{p-n} junctions seems to be sharp: for $\beta>2.5$ the
clean limit applies, for $\beta<0.7$ the disordered limit applies
and in between neither limit is valid.~\cite{footnote_fluc}.

Being sharply dependent on angle of incidence, transport through
potential steps in graphene should be sensitive to the presence of a
magnetic field, which bends electron trajectories. 
For $n_{\mathrm{bg}}=-n_{\mathrm{tg}}$ the predicted interface conductance in the clean limit is
\begin{equation}
G_{\mathrm{pn}}(B)=G_{\mathrm{pn}}(0)(1-(B/B_{\star})^{2})^{3/4},
\label{G_B}
\end{equation}
where $G_{\mathrm{pn}}(0)$ is the conductance at zero field, $B_{\star}= \hbar(e l)^{-1}
\sqrt{\pi \Delta n_{\mathrm{tg}}} $ and $l$ is the distance over
which the potential rises, which is proportional to the thickness
$d$ of the oxide~\cite{Levitov}.
We measure $R_{\mathrm{odd}}^{-1}$ as a function of magnetic field $B$ in two devices C540 and C1700 on the same
graphene sheet but with different top gate dielectric thickness $d$
(Table~1). We use the experimental $G_{\mathrm{pn}}(0)$ and
the best parameter $l$ to fit all curves within the same device (see Supporting material). The parameters $l$ for C540 and C1700
are found to be 65~nm and 55~nm respectively, whereas C1700 has the
thicker dielectric (see Table~\ref{Table1}). Further
theoretical work is needed to explain this discrepancy.

In conclusion, we show evidence for Klein tunneling across
potential steps in graphene with a quantitative agreement to a model with one 
free parameter describing screening properties in graphene. 
The crossover between clean and disordered regimes occurs as a function of the parameter $\beta$
around 1 as predicted by Fogler \emph{et al.}
\cite{fogler_effect_2008}. 
More work is needed to go into the fully ballistic regime, and also to measure directly the angle
dependence of Klein tunneling~\cite{cheianov_selective_2006}.

We thank J. A. Sulpizio for help with fabrication and
characterization, and M. Fogler, D. Novikov,  L. Levitov, and A. Young for enlightening discussions.  We also thank A. Savchenko for pointing out the need to take into account the resistance of the monopolar junction in the predictions for $R_{\mathrm{odd}}$.
While this work was under review, we became aware of related work by A. Young \emph{et al.}, 
in which evidence is seen for ballistic transport across a full npn junction~\cite{Young}.
This work was supported by the MARCO/FENA
program and the Office of Naval Research contract N00014-02-1-0986.
N. Stander was supported by a William R. and Sara Hart Kimball
Stanford Graduate Grant. Work was performed in part at the
Stanford Nanofabrication Facility of NNIN supported by the National
Science Foundation under Grant ECS-9731293. Critical equipment
(SEM,AFM) was obtained partly on Air Force Grants FA9550-04-1-0384 and
F49620-03-1-0256.

\newpage

{\LARGE Supplementary material}

section{Graphene characterization}

\begin{figure}[hptb]
\begin{center}
\includegraphics[width=8cm]{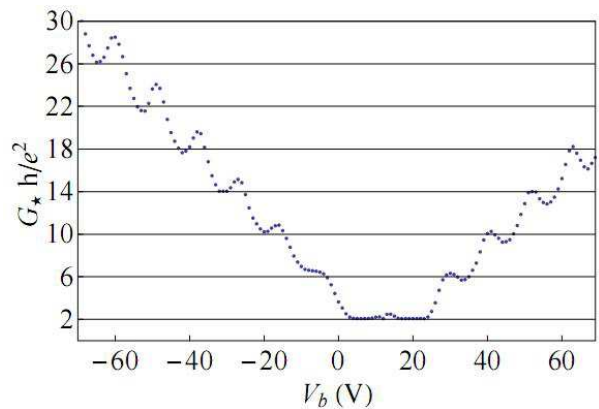}
\caption{2-probe conductance corrected using the estimated contact
resistance for sample C540 at a magnetic field of 8~T and a
temperature 4~K. \label{QHE}}
\end{center}\end{figure}

We measure 2-probe conductance in each sample at high magnetic field
(8~T), in order to verify it has the unique behavior of a single
sheet. For example Fig.~\ref{QHE} shows conductance $G_\star$
measured in sample C540 at 8~T (note that an estimated contact
resistance  $R_\mathrm{con}=1830~\Omega$ has been taken into account
$G_\star=(G^{-1}-R_\mathrm{con})^{-1}$). The plateaus in $G_\star$
are at values $2e^2/h$, $6e^2/h$ \ldots , characteristic of a single
layer. Appearance of peaks between plateaus was predicted by Abanin and Levitov, for a 2-probe measurement.~\cite{abaninlevitov}.

\begin{figure}[hbtp]
\begin{center}
\begin{minipage}{.3\linewidth}
\includegraphics[width=5cm]{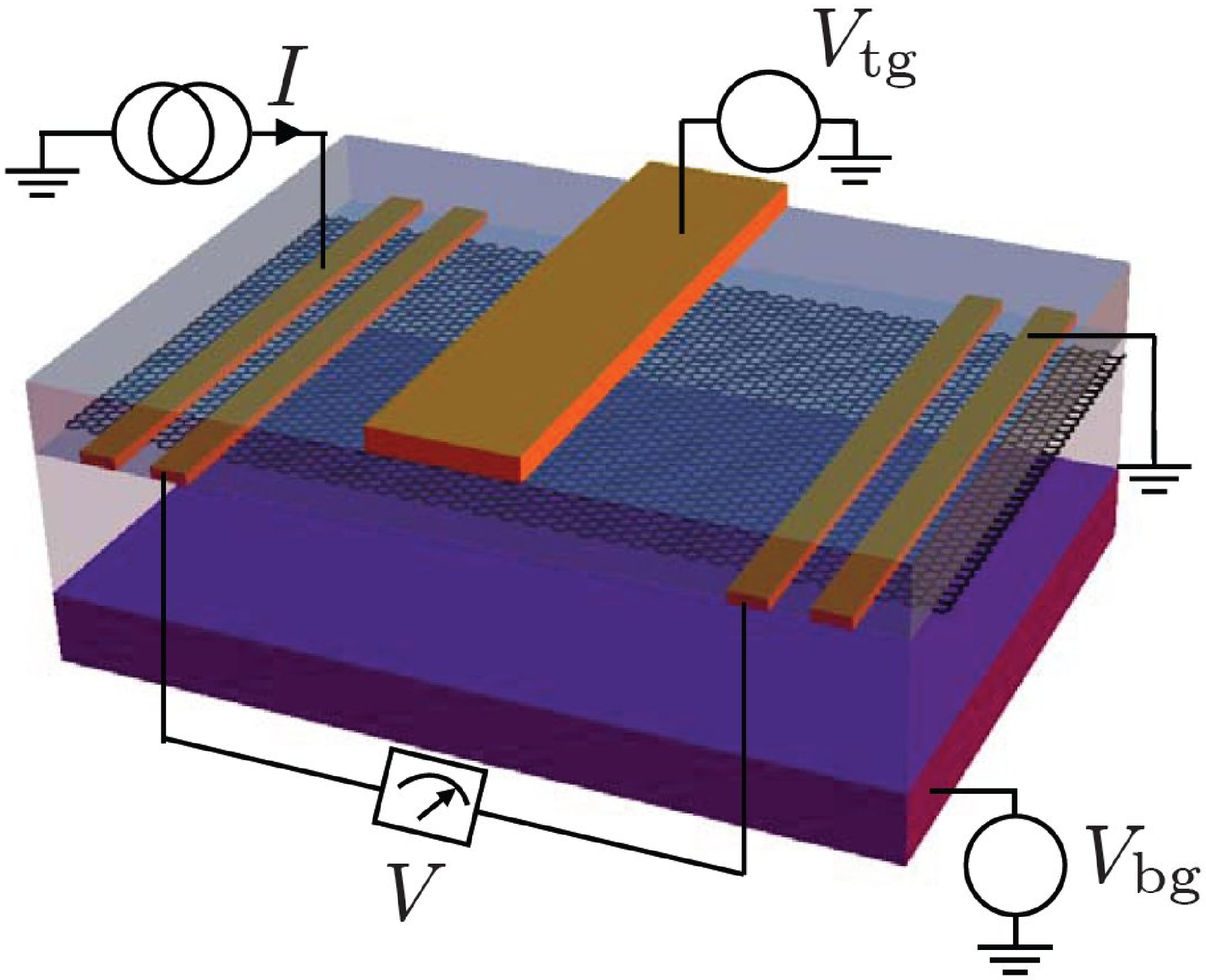}
\end{minipage}
\begin{minipage}{.3\linewidth}
\includegraphics[width=5cm]{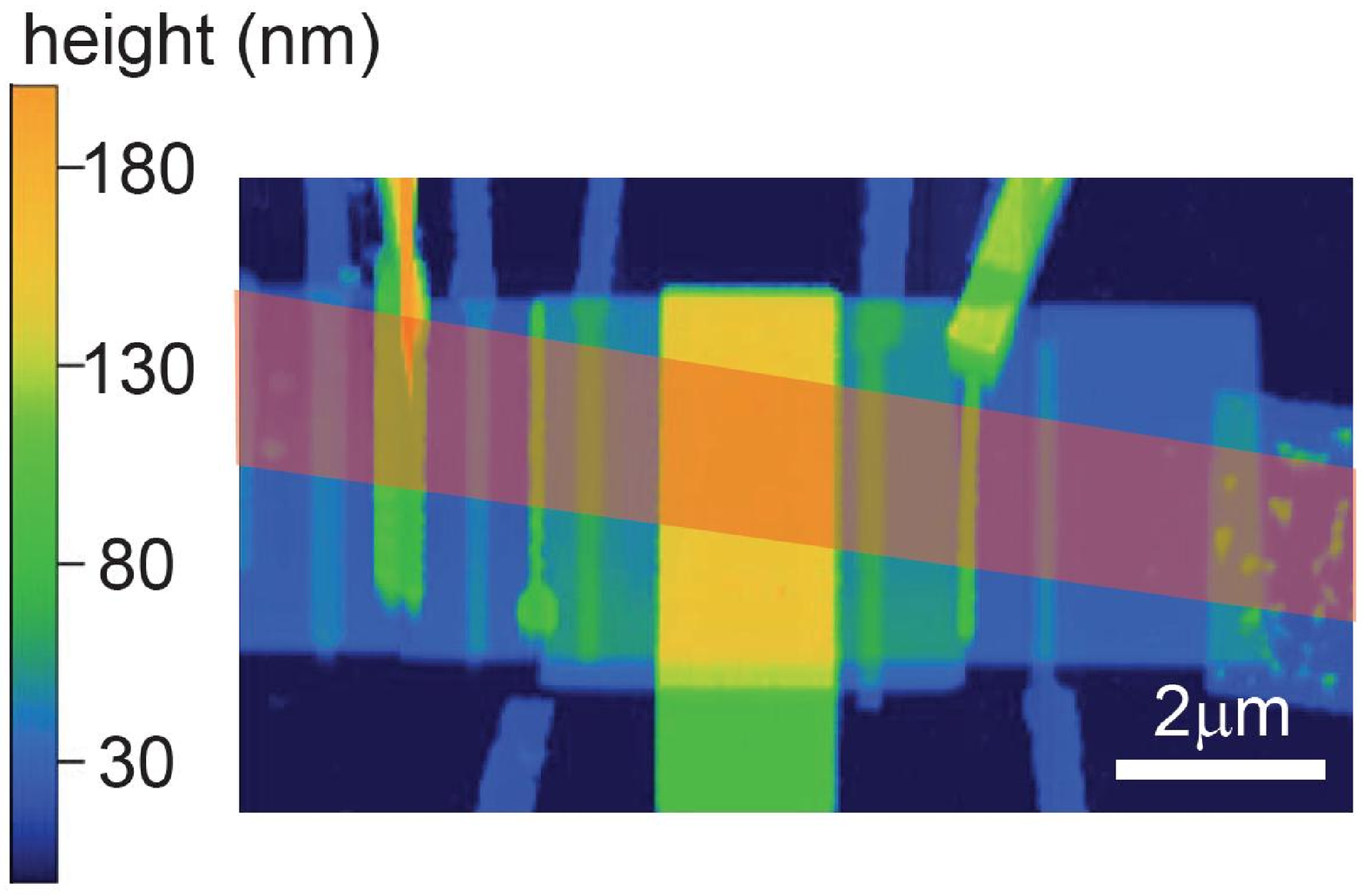}
\end{minipage}
\begin{minipage}{.3\linewidth}
\includegraphics[width=6cm]{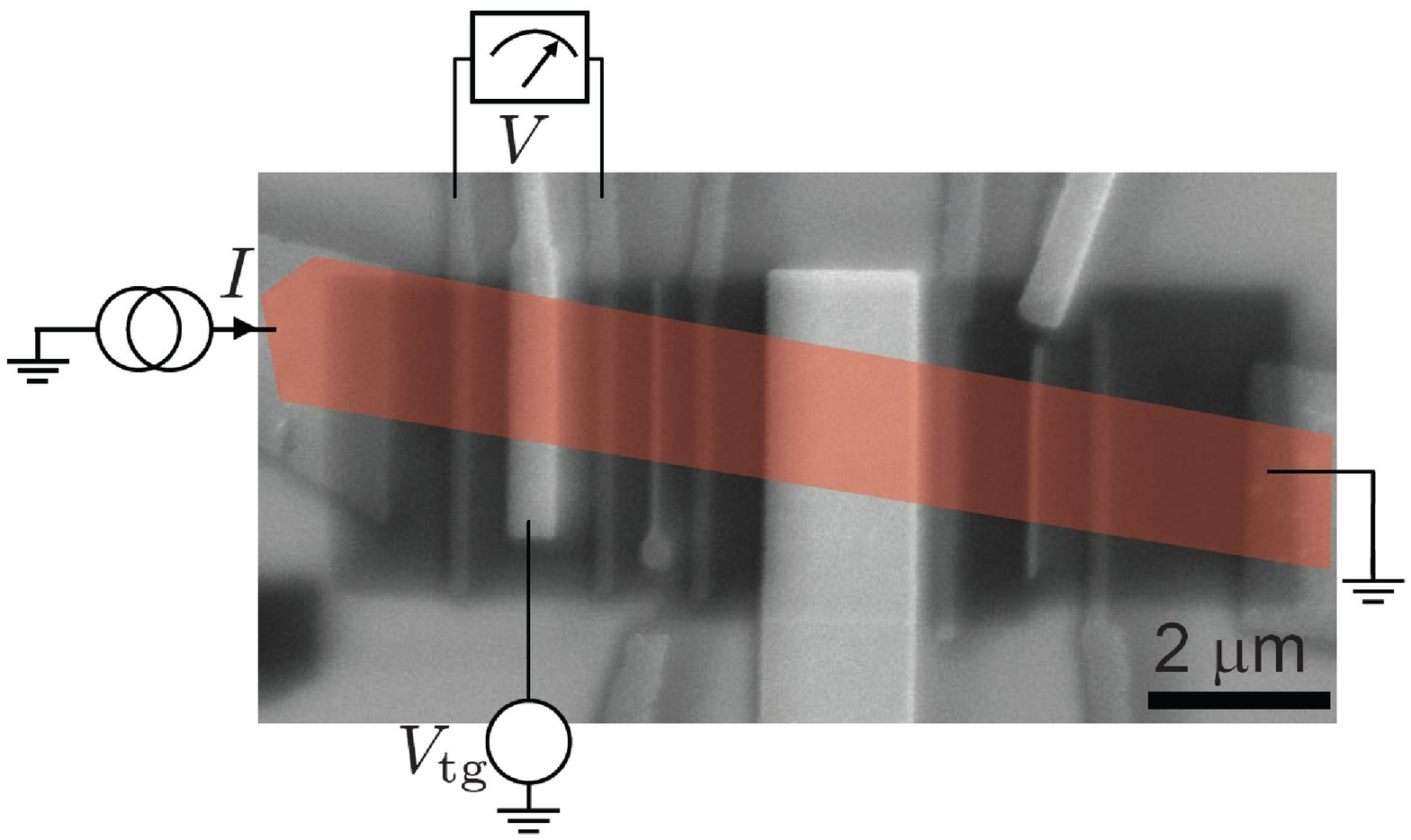}
\end{minipage}
\caption{Left: 3-dimensional schematic of representative device. Middle:
Atomic Force Microscope topograph of devices C540 and C1700. Right: Scanning Electron Microscope image with 4-probe measurement scheme for C540. \label{images}}
\end{center}
\end{figure}

\section{Extracting the odd part of the resistance}

Extracting the odd part of the resistance requires the
determination of three quantities: The ratio between the top gate capacitance $C_\mathrm{tg}$ and
the back gate capacitance $C_\mathrm{bg}$, $V_\mathrm{bg}^0$, which corresponds to zero average density far from the top gated region,and $V_\mathrm{tg}^0$ which corresponds to zero average density below the top gated region, when $V_\mathrm{bg}=V_\mathrm{bg}^0$. A good approximation to these parameters can be extracted from Fig.~1a of the paper, since the voltage offsets are the
coordinates of the global maximum in resistance and the slope of the
diagonal peaked line gives the ratio $C_\mathrm{tg}/C_\mathrm{bg}$.
However, the odd part turns out to be particularly sensitive to $C_\mathrm{tg}/C_\mathrm{bg}$,
so that a mere estimation of the peak position is not enough.

Instead, we measure the resistance for each device as a
function of voltages $V_\mathrm{bg}$ and $V_\mathrm{tg}$ at 8~T in
the Quantum Hall regime. The position of the transition between the
first and second conductance plateaus in $V_\mathrm{tg}$ for each
value of $V_\mathrm{bg}$ leads to a determination of the ratio
$C_\mathrm{tg}/C_\mathrm{bg}$ within $1\%$. The determination of the voltages
$V_\mathrm{bg}^0$ and $V_\mathrm{tg}^0$ can be done accurately by
symmetrizing the resistance in Fig.~1 of the paper with respect to
the point $(V_\mathrm{bg}^0,V_\mathrm{tg}^0)$:\be
R(V_\mathrm{bg}^0+\Delta V_\mathrm{bg},V_\mathrm{tg}^0+\Delta
V_\mathrm{tg})\leftarrow R(V_\mathrm{bg}^0-\Delta
V_\mathrm{bg},V_\mathrm{tg}^0-\Delta V_\mathrm{tg})
\ende
and choosing the point $(V_\mathrm{bg}^0,V_\mathrm{tg}^0)$ which leaves
this resistance the most unchanged.

\begin{table}[hbtp]
\begin{center}

\begin{tabular}{|c|c|c|c|c|c|c|c|c|}\hline
Sample & L (nm) & w ($\mu$m) & d (nm) & $V_\mathrm{bg}^0$~(V) &
$V_\mathrm{tg}^0$~(V) & $C_\mathrm{tg}$ (nF.cm$^{-2}$) &
$\mu$~(cm$^2$V$^{-1}$s$^{-1}$) & $\left\langle \beta
\right\rangle$
\\\hline
A60 & 60 & 4.3 & 34 & 25.65 & -1.36 & 92 & 1800 & 7.6\\
B100 & 100 & 2.1 & 42 & 9.35 & -0.49 & 69 & 1700 & 3.8\\
B220 & 220 & 2.1  & 42 & 10.95 & -0.73 & 69 & 1700 & 3.5\\
C540 & 540 & 1.74  & 25 & 18.65 & -2.42 & 107 & 1400 & 7.9\\
A860 & 860 & 3.6 & 34 & 25.5 & -2.35 & 92 & 1800 & 7.9\\
C1700 & 1700 & 1.74 & 47 & 13.4 & -1.35 & 52 & 1300 & 1.9\\
\hline
\end{tabular}
\caption{Geometrical properties of the samples: $L$- top gate
length, $w$- interface width, and $d$- top gate dielectric
thickness. Same letter for two devices indicates same graphene
sheet. All dimensions were taken by Scanning Electron Microscope
(SEM) and Atomic Force Microscope (AFM) images. The gate voltage
offsets $V_\mathrm{bg}^0$ and $V_\mathrm{tg}^0$ and the capacitance
of the top gate determined from the procedure described in the text
are reported here. The mobility $\mu$ is estimated from the slope at
the origin of the conductance measured as a function of back gate
voltage. These low values are due to the PMMA cross-linking step.
\label{Table1}}
\end{center}
\end{table}

\begin{figure}[hptb]
\begin{center}
\includegraphics[width=8cm]{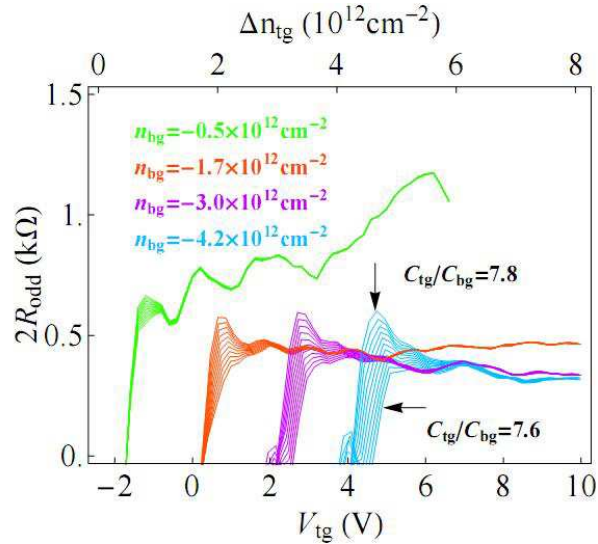}
\caption{$2R_{\mathrm{odd}}$ as a function of $V_{\mathrm{tg}}$ for several $n_{\mathrm{bg}}$ shown at the upper left corner for sample C540. For each chosen $n_{\mathrm{bg}}$, we plot the corresponding curves with $C_{tg}/C_{bg}$ between $7.6$ and $7.8$ in steps of 0.02. This reflects a spread of $\pm 1\%$ from the value of $7.7$, which we use in the paper. \label{capa}}
\end{center}
\end{figure}

Still, the uncertainty of $1\%$ on the capacitance ratio $C_\mathrm{tg}/C_\mathrm{bg}$
leads to some uncertainty on the odd part $R_\mathrm{odd}$ of the
resistance. However, this uncertainty remains negligible except at low densities
$n_\mathrm{tg}$ (see Fig.~\ref{capa}).

\section{Comparing the experimental value $R_{\mathrm{odd}}$ to theoretical models of the junction interface resistance}

\begin{figure}[hptb]
\begin{center}
\hspace{-3cm}
\begin{minipage}{.3\linewidth}
\includegraphics[width=8cm]{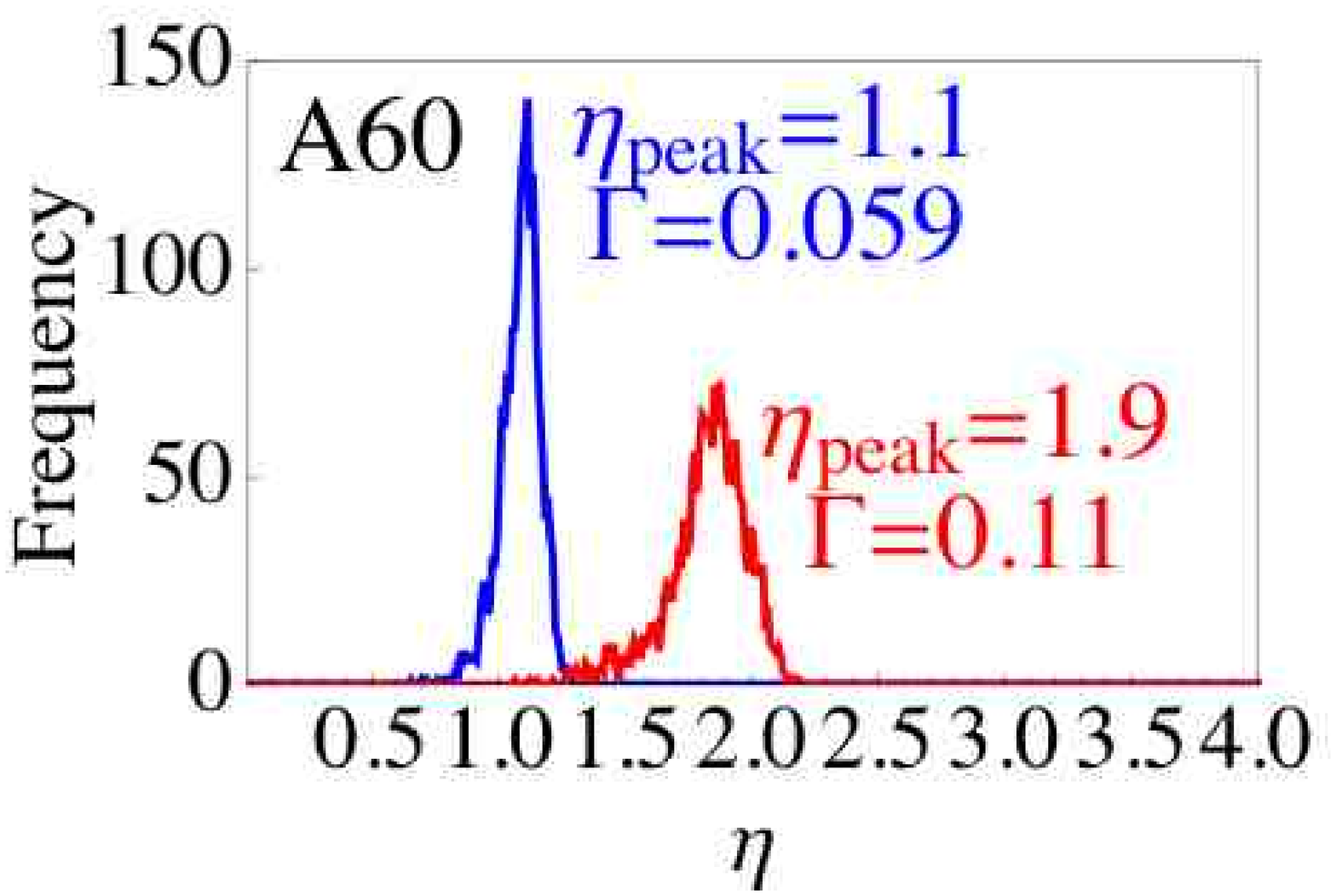}
\end{minipage} \hspace{3cm}
\begin{minipage}{.3\linewidth}
\includegraphics[width=8cm]{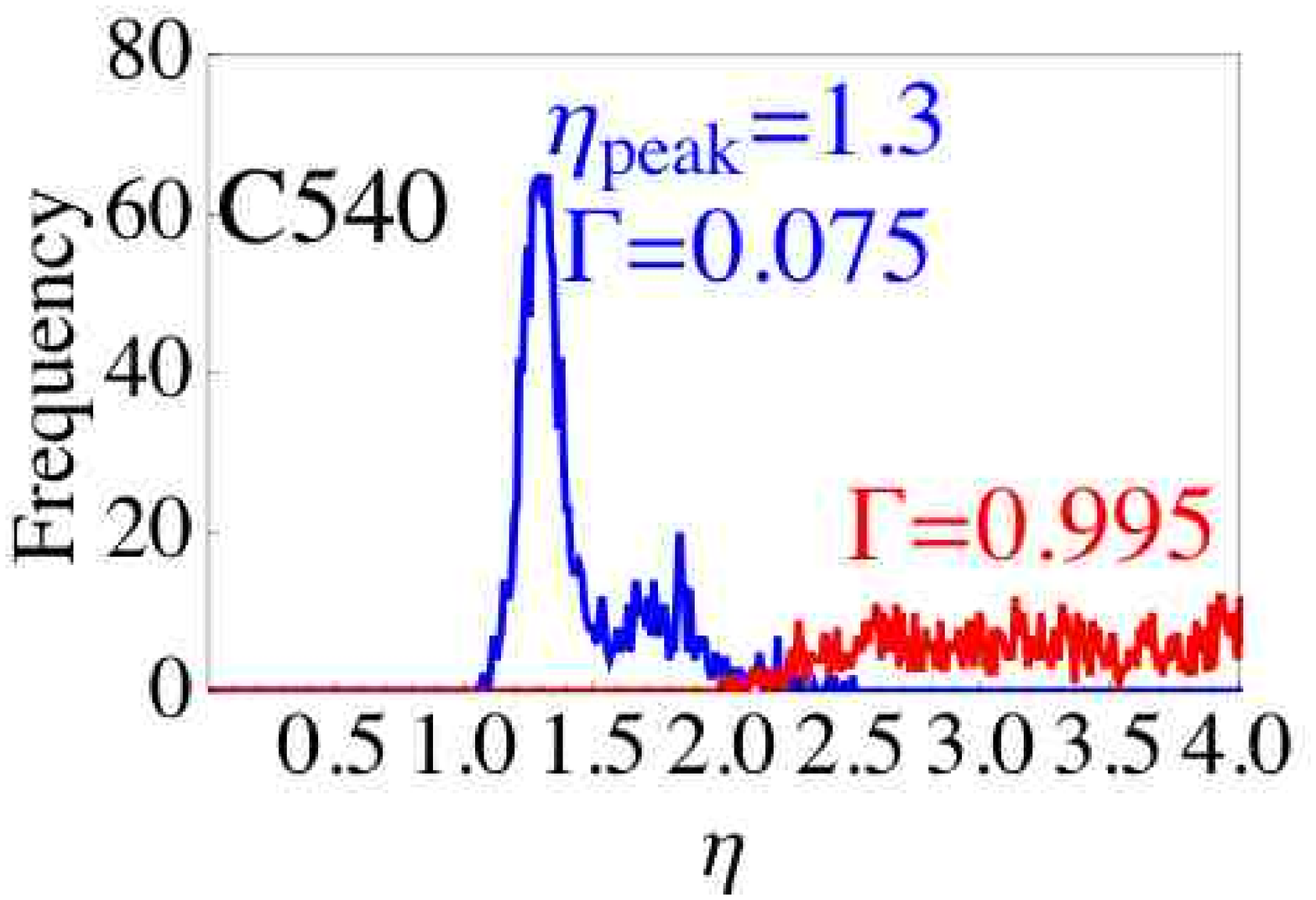}
\end{minipage}
\caption{\textbf{a)} Histograms of $\eta$ using a diffusive model (red)
or a ballistic model (blue) for device A60. The peak value
$\eta_{\mathrm{peak}}$ and peak width
($2\Gamma$) shown in the figure were taken from
a lorentzian fit Eq.~(\ref{fit}) to each theory. The histogram bins are 0.01 wide.
\textbf{b)} Same as a) for device C540. Using the diffusive theory for C540, we could not fit properly $\eta$ thus we report the standard deviation in eta as the value of $\Gamma$.\label{Statistics}}
\end{center}
\end{figure}

\begin{figure}[hptb]
\begin{center}
\includegraphics[width=8cm]{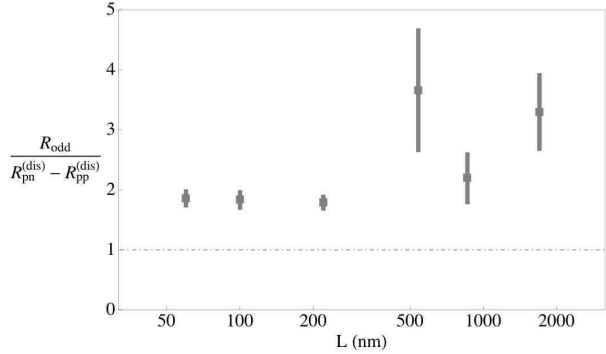}
\caption{The vertical lines show the spread of
$\eta$ when using diffusive model (Eq.~(2) in the main paper) for
densities such that~$\left|n_{\mathrm{bg}}\right|$,$\left|n_{\mathrm{tg}}\right|>10^{12}\mathrm{cm}^{-2}$.
the lines are centered on the average value of the
histogram.\label{dif}}
\end{center}
\end{figure}

Figures~3a and 3b in the paper show the experimental
$R_{\mathrm{odd}}$ in comparison to the theoretical
$R_{\mathrm{np}}-R_{\mathrm{pp}}$, for the two devices A60
and C540 at several $n_{\mathrm{bg}}$ within both clean and disordered models. In order to quantify the
compatibility of the theory to the experiment, we define
$\eta(V_{\mathrm{bg}},V_{\mathrm{tg}})$ as the ratio
$R_{\mathrm{odd}}/(R_{\mathrm{np}}-R_{\mathrm{pp}})$. We determine
the ratio $\eta$ for all measured densities
$|n_\mathrm{bg}|,n_\mathrm{tg}>10^{12}\mathrm{cm}^{-2}$ and
calculate the corresponding histogram for $\eta$ using two models of transmission across a single potential step in graphene: diffusive and ballistic. $\eta=1$
corresponds to perfect agreement between theory and experiment in
the limit where $L\gg l_e$ (see paper). Figures ~\ref{Statistics}a
and Figs.~\ref{Statistics}b show two histograms of
$\eta$ each, for devices A60
and C540, respectively. The red histogram is using a diffusive model while the blue one is using a ballistic model. We follow the same procedure for all
devices, and extract the value $\eta$ associated with the the diffusive theory and ballistic theory, at the
center of the peaked histogram together with its width
($2\Gamma$) , by fitting the data to the following
Lorentzian:
\begin{equation}
\mathrm{Frequency}=\frac{a}{(\eta-\eta_{\mathrm{peak}})^{2}+(\Gamma)^{2}}
\label{fit}
\end{equation}
In the paper, Fig.~4 presents
$\eta_{\mathrm{peak}}$ and small error bars for
$\eta$ when using the ballistic model with a fitting parameter $c_1=1.35$. This is complemented here by
Fig.~\ref{dif} showing the wide spread of $\eta$ for some devices, along with a much lower predicted value of $R_{\mathrm{np}}-R_{\mathrm{pp}}$  when using the diffusive model. Note that B100 and
B220 have a relatively smaller spread in $\eta$ when using the diffusive model, which is due to a smaller
range of densities. In contrast the relatively small spread of
$\eta$ when using the diffusive theory for A60 is not due to a smaller range of
densities nor due to the short dimension of the top gate. We
currently do not understand this feature, although the spread is still larger than the spread in $\eta$ when using the ballistic model. 

\section{Multiple reflections between interfaces of a potential barrier}

\label{multiplerefl}

One of the goals of the main paper was to investigate the transition
from diffusive to ballistic transport through the potential barrier by making
the top gate length smaller than the mean free path of the carriers.
In this limit the transport across the whole potential barrier
is expected to be ballistic (no disorder), and charge carriers are subject to multiple
reflections on the two interfaces of the barrier.

\subsection{Without phase coherence}

The transmission probability $T_\Sigma$ across the whole potential
barrier is related to the transmission probability $T$ across a
single interface by: \be T_\Sigma=T(T+(1-T)^2(T+(1-T)^2(T\ldots
\ende
Hence, \be T_\Sigma=\frac{1}{2T^{-1}-1}.
\ende
Therefore the total conductance for a width $w$ is given by \be
G_\Sigma=\frac{4e^2}{h}\frac{w}{2\pi}\int_{-\infty}^{+\infty}T_\Sigma(k_y)\mathrm{d}k_y
\ende
where $k_y$ is the component of the wavevector $\vec{k}$ along the
potential interface. One can compare this to the conductance across
a single interface \be
G=\frac{4e^2}{h}\frac{w}{2\pi}\int_{-\infty}^{+\infty}T(k_y)\mathrm{d}k_y
\ende
so that \be
G_\Sigma/G=\frac{\int_{-\infty}^{+\infty}(2T(k_y)^{-1}-1)^{-1}\mathrm{d}k_y}{\int_{-\infty}^{+\infty}T(k_y)\mathrm{d}k_y}
\ende
According to Ref.~\cite{cheianov_selective_2006}, $T(k_y)=e^{-\gamma
k_y^2}$ with $\gamma>0$ therefore,
 \be
G_\Sigma/G=\sum_{n=1}^\infty\frac{1}{2^k\sqrt{k}}\approx 0.81
\ende

Using the notations of our paper, this translates into a resistance
\be R_\Sigma\approx 1.24 R_\mathrm{np}^{(\mathrm{bal})}.
\ende
Therefore, for a length $L$ small enough ($L\ll l_e$), the odd part
of the resistance should be such that \be 2R_\mathrm{odd}\approx
1.24 R_\mathrm{np}^{(\mathrm{bal})}.
\ende

As seen from Fig.~4 in the paper, this regime is never achieved
fully in the experiments but may be the cause of the smaller $\eta$
for the shortest top gate.

\subsection{Including phase coherence}

In the phase coherent regime, the above derivation remains valid up
to a phase term in the transmission: \be
T_\Sigma=|T(T+(1-T)^2\mathrm{e}^{i E \Delta
t/\hbar}(T+(1-T)^2\mathrm{e}^{i E \Delta t/\hbar}(T\ldots|
\ende where $\Delta t= 2L\cos\theta$ is the time spent between back and forth bounces and $\theta$ is the angle of incidence. This simplifies into\be T_\Sigma=T^2\left|1-(1-T)^2\exp\left[i2\pi n L(\pi
n-k_y^2)^{-1/2}\right]\right|^{-1}
\ende with $n$ the density below the top gate. Phase coherent length in our devices is of the order of a few microns, extracted from a similar device in Ref.~\cite{huard_transport_2007}


\section{\emph{n-p} junctions in finite magnetic field}

\begin{figure}[hbtp]
\begin{center}
\begin{minipage}{.49\linewidth}
\includegraphics[width=6cm]{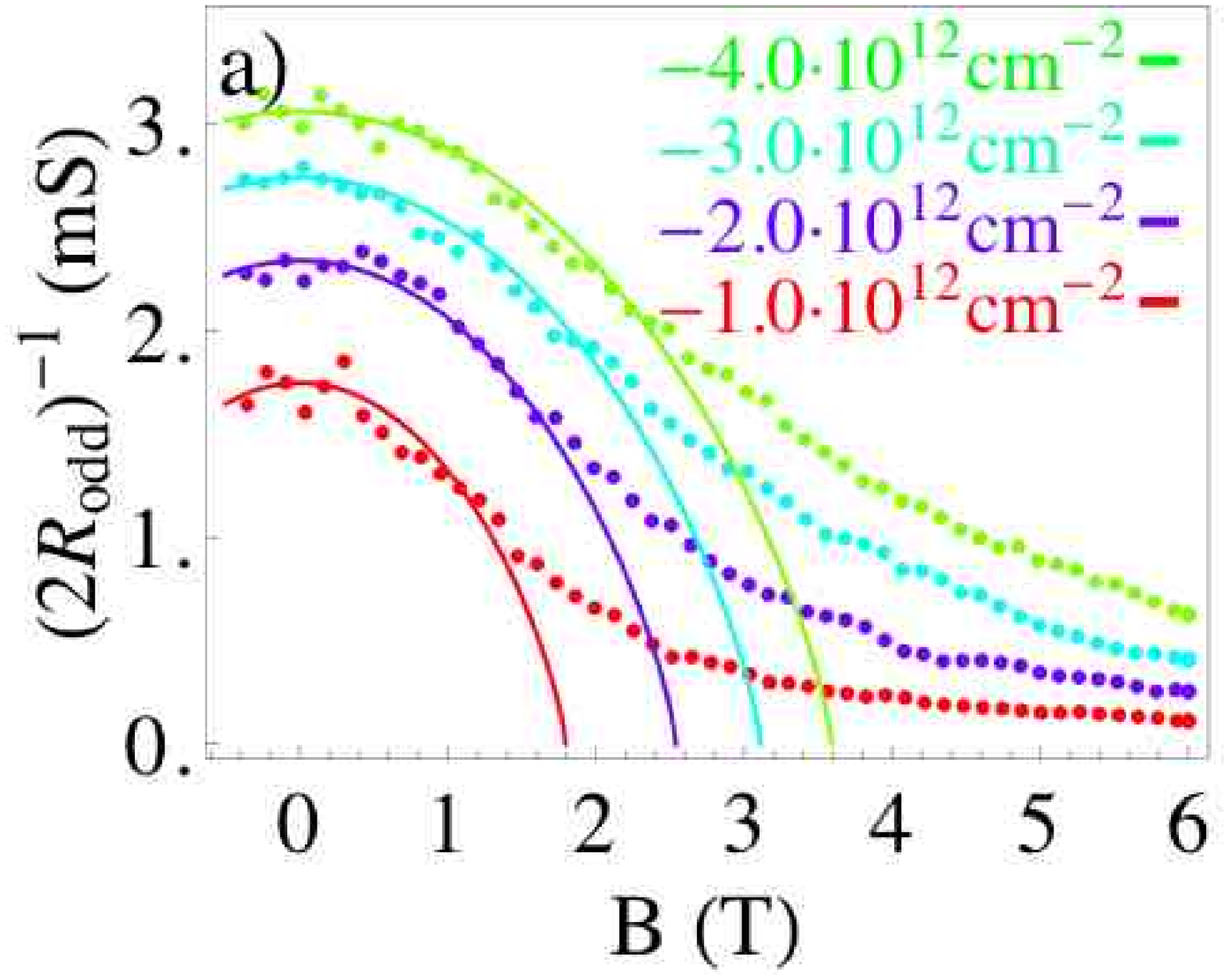}
\end{minipage} \hfill
\begin{minipage}{.49\linewidth}
\includegraphics[width=6cm]{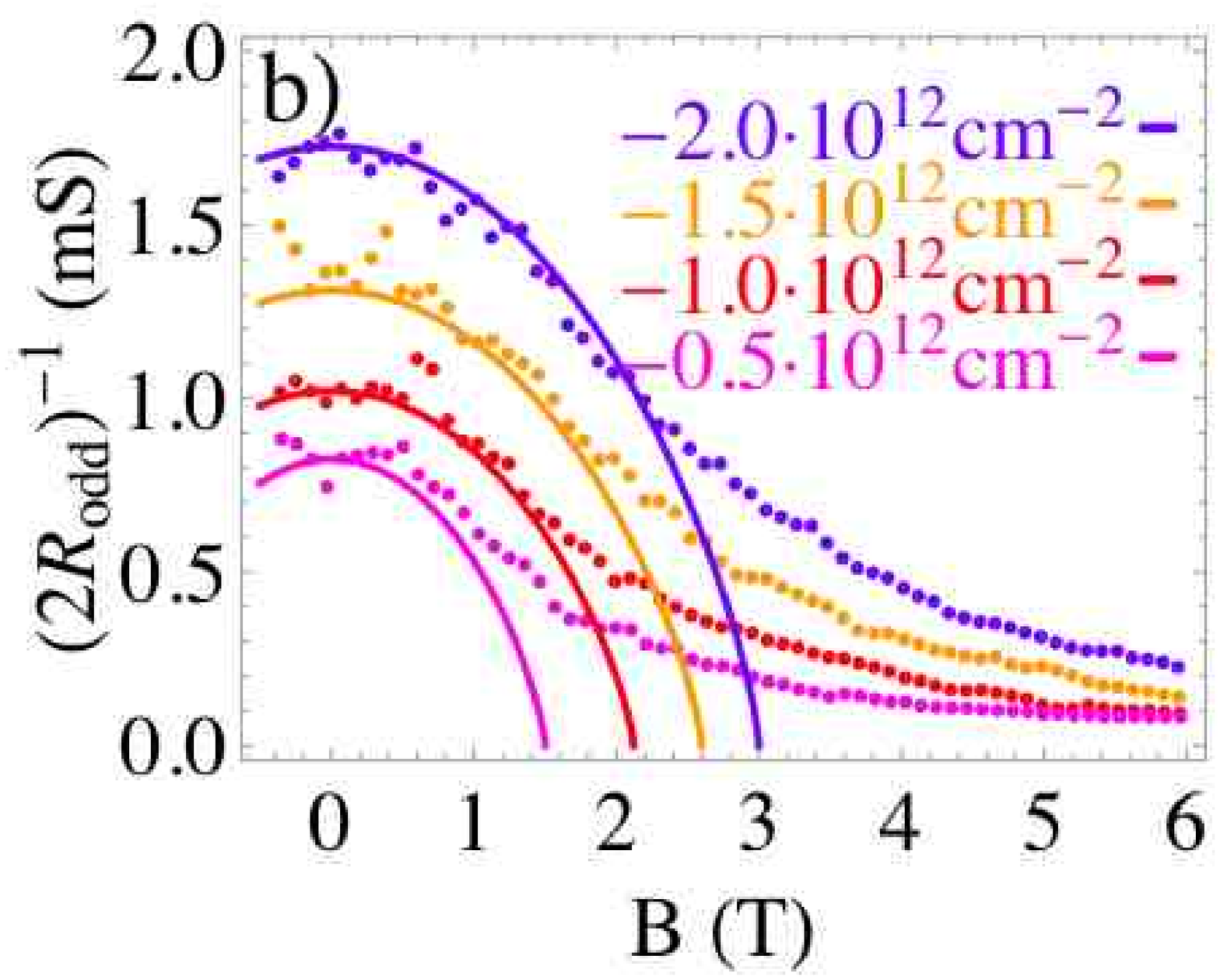}
\end{minipage}
\caption{\textbf{a)} $(R_{\mathrm{odd}})^{-1}$ for device C540 as a function of magnetic field $B$ for several density profiles with $n_{\mathrm{bg}}=-n_{\mathrm{tg}}$ ($n_{\mathrm{bg}}$ is labeled).  The theoretical curves using Eq.~(\ref{G_B}) (solid lines) are fitted with $l=65~\mathrm{nm}$ to the experimental curves (dots).\textbf{b)} Same as a) for device C1700. The fitting parameter used was $l=55~\mathrm{nm}$. \label{Figure5}}
\end{center}
\end{figure}

As explained in the paper transport through
potential steps in graphene should be sensitive to the presence of a
magnetic field, which bends electron trajectories. 
For instance, in the clean limit the angle at which carriers are transmitted perfectly should be
given by $\arcsin(B/B_{\star})$ where $B_{\star}= \hbar(e l)^{-1}
\sqrt{\pi \Delta n_{\mathrm{tg}}} $ and $l$ is the distance over
which the potential rises, which is proportional to the thickness
$d$ of the oxide~\cite{Levitov}. For $n_{\mathrm{bg}}=-n_{\mathrm{tg}}$ the
predicted interface conductance is
\begin{equation}
G_{\mathrm{pn}}(B)=G_{\mathrm{pn}}(0)(1-(B/B_{\star})^{2})^{3/4},
\label{G_B}
\end{equation}
where $G_{\mathrm{pn}}(0)$ is the conductance at zero field. 
Since Eq.~(\ref{G_B}) is a prediction for the conductance of a single \emph{p-n} interface and $l_{e}\ll L$ in
both devices, $R_{\mathrm{odd}}^{-1}$ can be interpreted as the
conductance of a single \emph{p-n} interface (canceling out the monopolar bulk magnetoresistance, whose source in not well understood). For several gate voltages such that $n_{\mathrm{bg}}=-n_{\mathrm{tg}}$, we measure
$R_{\mathrm{odd}}^{-1}$ as a function of magnetic field $B$
(Fig.~\ref{Figure5}) in two devices C540 and C1700 on the same
graphene sheet but with different top gate dielectric thickness $d$
(Table~1). 
We use the experimental $G_{\mathrm{pn}}(0)$ and
the best parameter $l$ to fit all curves within the same device. The parameters $l$ for C540 and C1700
are found to be 65~nm and 55~nm respectively, whereas C1700 has the
thicker dielectric (see Table~\ref{Table1}). 

We also show here how to extract the \emph{n-p} interface
conductance in the presence of magnetic field. Both C540 and C1700
satisfy the condition $l_{e}\ll L$ (Fig.~4 of our paper), thus the
barrier resistance can be viewed as that of two \emph{n-p} interfaces in
series. In this case
$(2R_{\mathrm{odd}})^{-1}=(R_{\mathrm{pnp}}-R_{\mathrm{ppp}})^{-1}$,
where $R_{\mathrm{pnp}}=G_{\mathrm{pnp}}^{-1}$
($R_{\mathrm{ppp}}=G_{\mathrm{ppp}}^{-1}$) is the resistance of the
barrier when $n_{bg}=-n_{tg}$ ($n_{bg}=n_{tg}$).
Figures~\ref{Bfield}a-d show $G_{\mathrm{pnp}}$ and
$G_{\mathrm{ppp}}$ for C540 and C1700, as a function of magnetic
field, at several $n_{bg}$. The flatness of the $n_{bg}=0$ curve is
a measure of how well $V_{\mathrm{bg}}^{0}$ and
$V_{\mathrm{tg}}^{0}$ were determined. Also,
$G_{\mathrm{ppp}}>G_{\mathrm{pnp}}$ at all measured densities, in
both devices, which is consistent with the zero magnetic field case.
Finally, we note a weak localization dip in both devices C540 and
C1700 in the conductance near $B=0$ for all densities.

\begin{figure}[hptb]
\begin{center}
\hspace{-3cm}
\begin{minipage}{.3\linewidth}
\includegraphics[width=6cm]{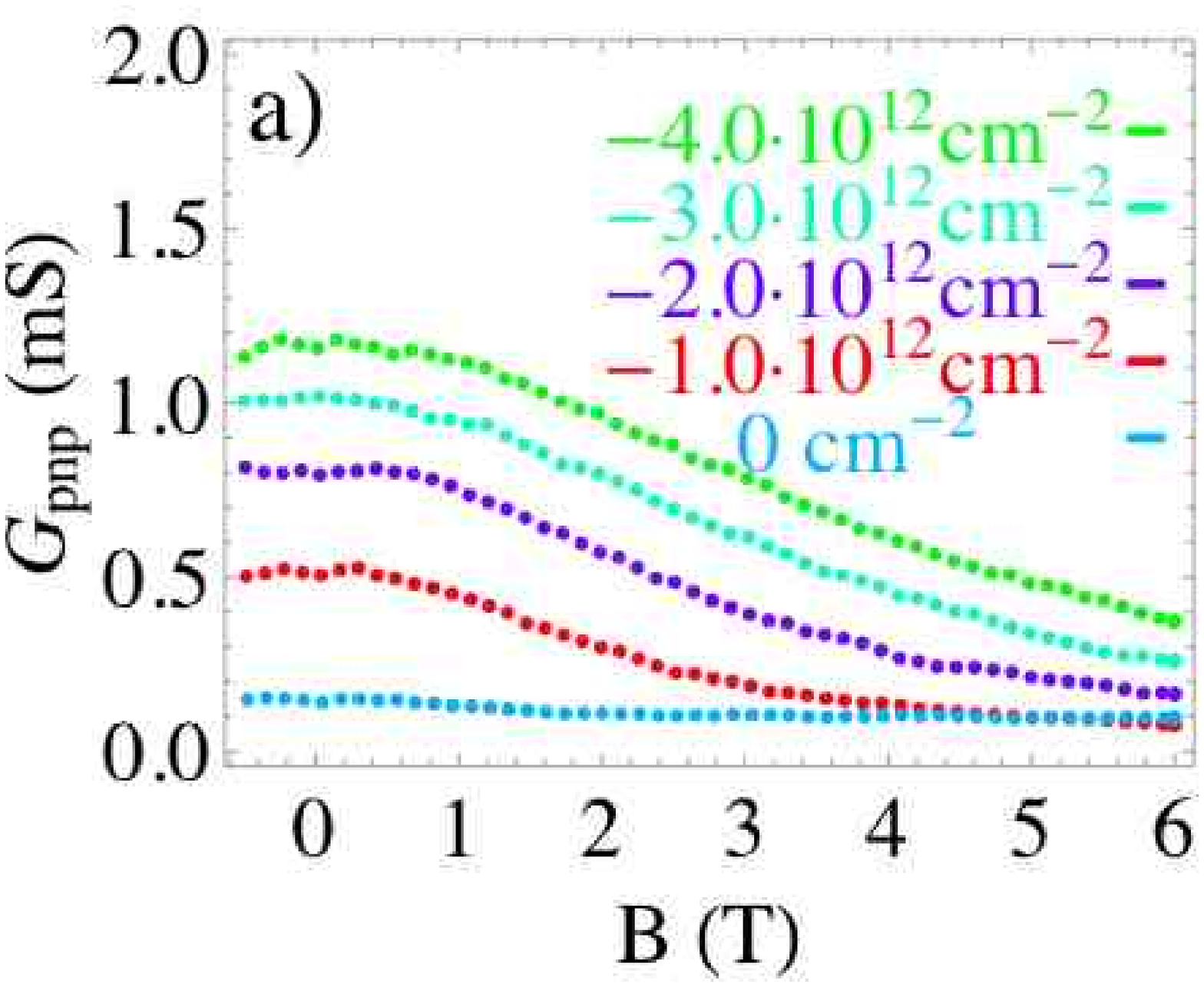}
\end{minipage} \hspace{2cm}
\begin{minipage}{.3\linewidth}
\includegraphics[width=6cm]{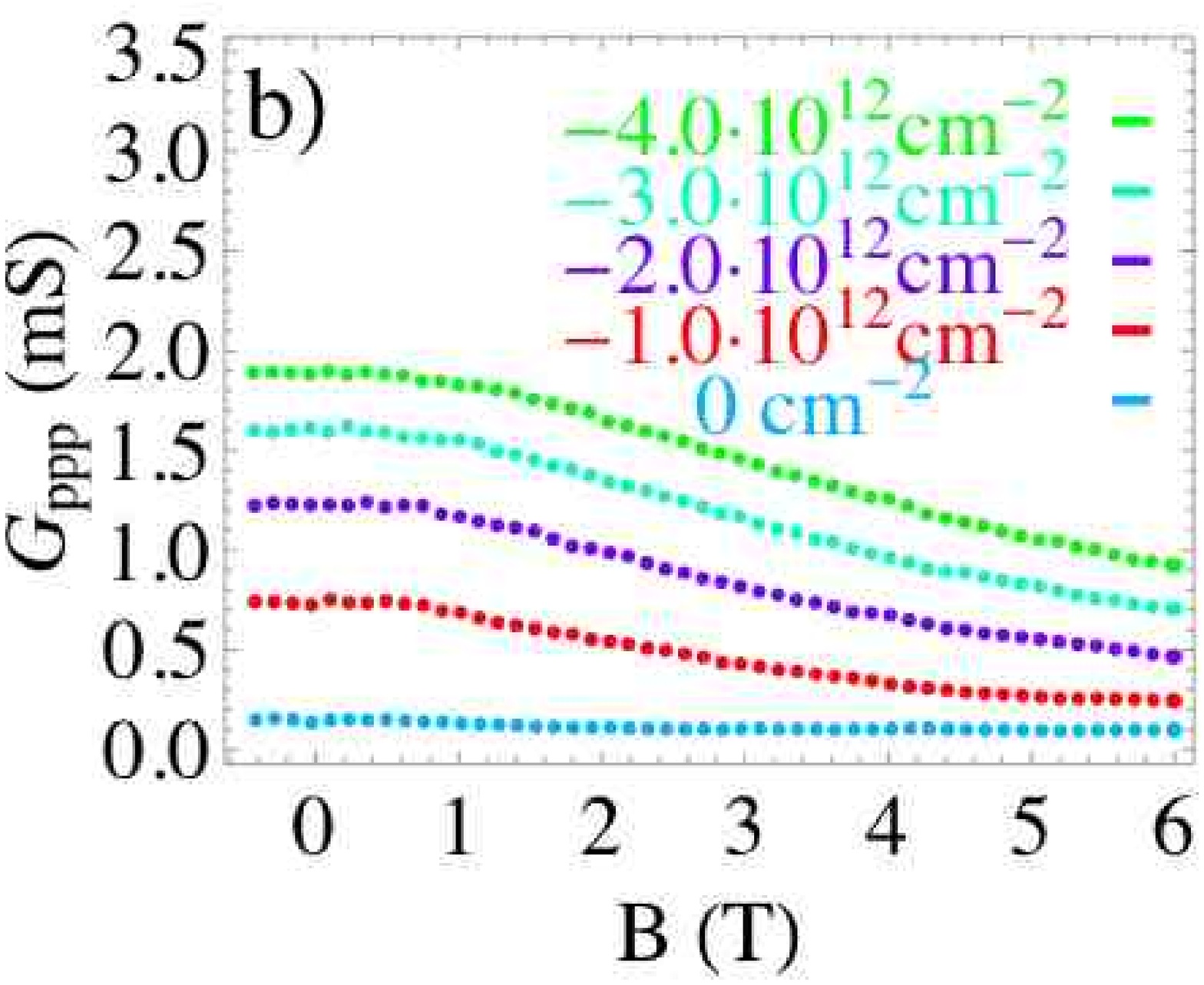}
\end{minipage}
\end{center}
\end{figure}

\begin{figure}[hptb]
\begin{center}
\hspace{-3cm}
\begin{minipage}{.3\linewidth}
\includegraphics[width=6cm]{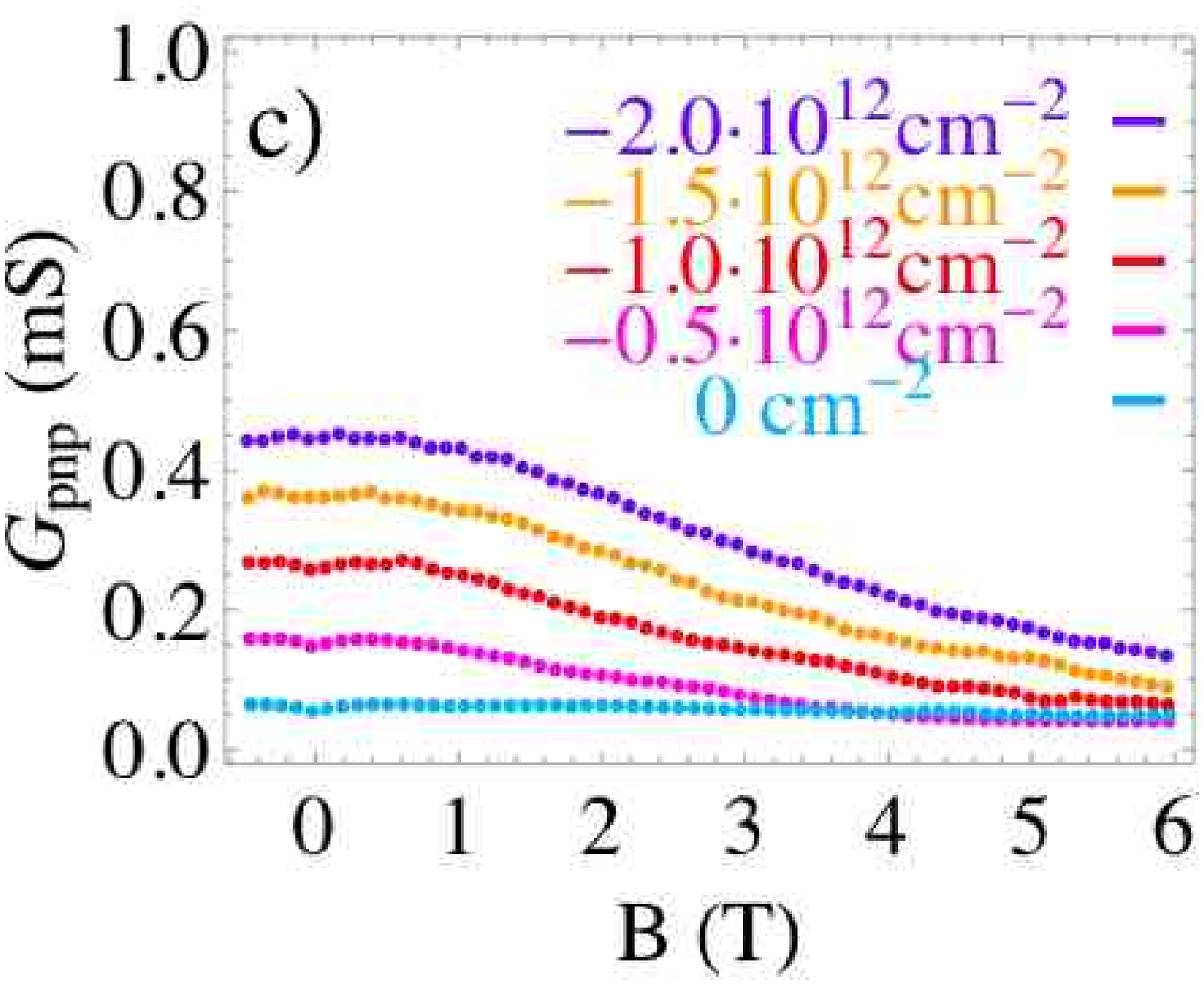}
\end{minipage} \hspace{3cm}
\begin{minipage}{.3\linewidth}
\includegraphics[width=6cm]{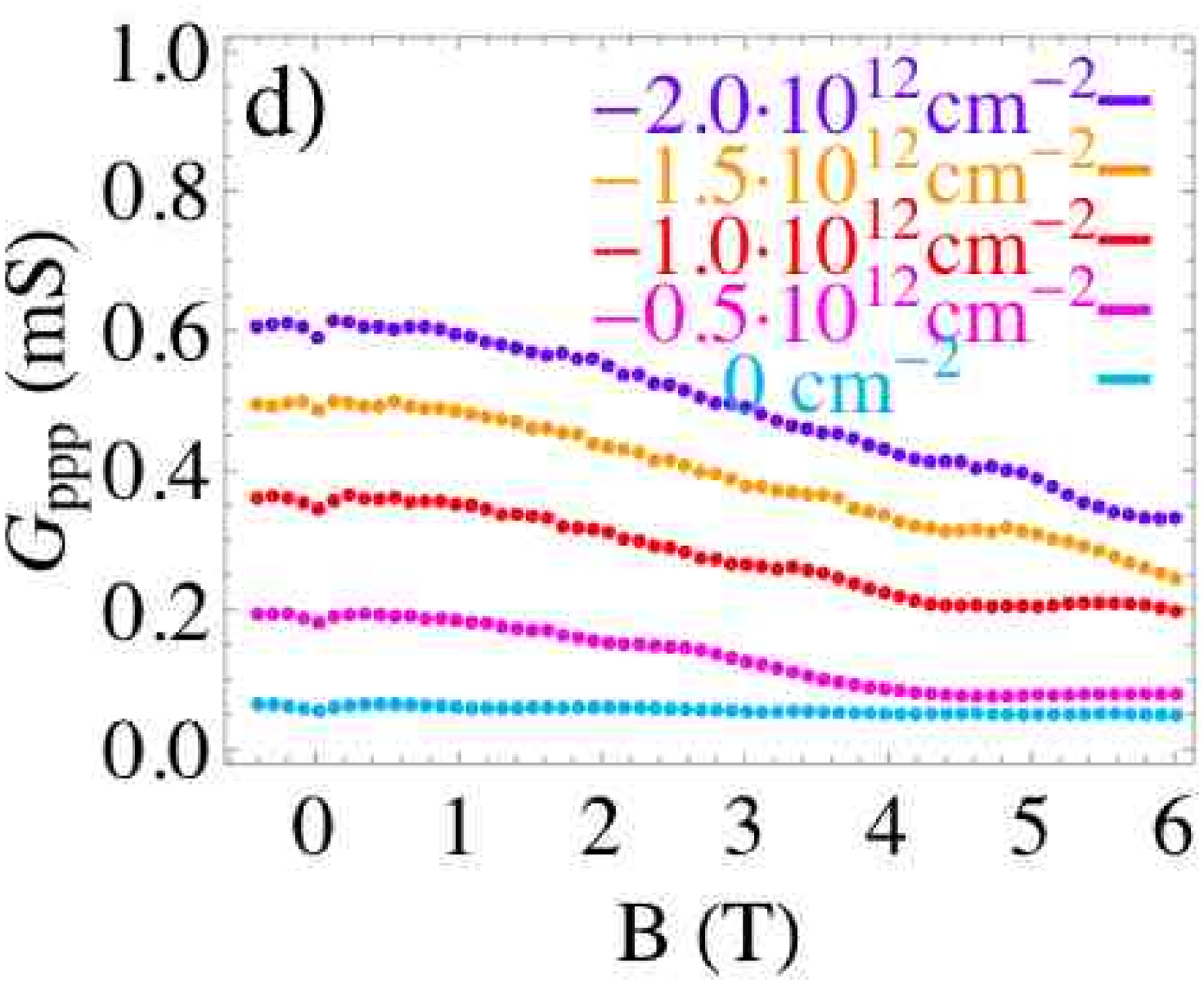}
\end{minipage}
\caption{\textbf{a)} $G_{\mathrm{pnp}}=R_{\mathrm{pnp}}^{-1}$ as a function of magnetic field for several $n_{bg}$, with $n_{tg}=-n_{bg}$, for device C540. $n_{bg}$ densities are presented on the right hand side of the figure. \textbf{b)} $R_{\mathrm{ppp}}^{-1}=G_{\mathrm{ppp}}$ as a function of magnetic field for several $n_{bg}$, with  $n_{tg}=n_{bg}$, for device C540. \textbf{c)}-\textbf{d)} Same as a) and b) for device C1700.  \label{Bfield}}
\end{center}
\end{figure}

\subsection{Fabrication details}

The substrate used in these experiments is a highly n-doped Si wafer
with a nominal resistivity of less than
$0.005~\Omega\cdot\mathrm{cm}$. Standard $1-10~\Omega \cdot
\mathrm{cm}$ wafers experience carrier freeze-out and hence
hysteretic response to applied gate voltage at temperatures below
$4\mathrm{K}$.

All graphene sheets were produced by successive mechanical
exfoliation of Highly Oriented Pyrolytic Graphite grade ZYA from
General Electric (distributed by SPI) using an adhesive tape (3M
Scotch Multitask tape with gloss finish), then deposited onto a
layer of SiO$_2$ 297~nm thick grown by dry oxidation at
$1500^{\circ}~\mathrm{C}$ on a highly n-doped Si substrate, which
serves as a global back gate. Before deposition of graphene, the
substrate was cleaned by Piranha etch. After suitable sheets were located with respect to alignment marks by optical microscopy, metallic probes were patterned using standard electron
beam lithography followed by electron beam evaporation of Ti/Au
(5~nm/25~nm thick). Afterward, the graphene sheets were etched in
dry oxygen plasma (1:9 O$_2$:Ar) into the desired shape, and one or
two layers of Polymethyl Methacrylate (PMMA, molecular mass 950K or 495K
at 2\% in anisole) were spun on top of it, then cross-linked using
$30~\mathrm{keV}$ electron beam with a dose of $2 \times 10^{4}~\mu
\mathrm{C.cm}^{-2}$. In a final e-beam lithography step, the top
gates were patterned on top of the cross-linked layer, followed by
electron beam evaporation of Ti/Au ($5$~nm/$45$~nm-55~nm thick).


\end{document}